*Research Article*

# Application of Multivariate Adaptive Regression Splines (MARSplines) for Predicting Hansen Solubility Parameters Based on 1D and 2D Molecular Descriptors Computed from SMILES String

**Maciej Przybyłek , Tomasz Jeliński , and Piotr Cysewski**

*Chair and Department of Physical Chemistry, Faculty of Pharmacy, Collegium Medicum of Bydgoszcz,
Nicolaus Copernicus University in Toruń, Kurpińskiego 5, 85-950 Bydgoszcz, Poland*

Correspondence should be addressed to Tomasz Jeliński; tomasz.jelinski@cm.umk.pl

A new method of Hansen solubility parameters (HSPs) prediction was developed by combining the multivariate adaptive regression splines (MARSplines) methodology with a simple multivariable regression involving 1D and 2D PaDEL molecular descriptors. In order to adopt the MARSplines approach to QSPR/QSAR problems, several optimization procedures were proposed and tested. The effectiveness of the obtained models was checked via standard QSPR/QSAR internal validation procedures provided by the QSARINS software and by predicting the solubility classification of polymers and drug-like solid solutes in collections of solvents. By utilizing information derived only from SMILES strings, the obtained models allow for computing all of the three Hansen solubility parameters including dispersion, polarization, and hydrogen bonding. Although several descriptors are required for proper parameters estimation, the proposed procedure is simple and straightforward and does not require a molecular geometry optimization. The obtained HSP values are highly correlated with experimental data, and their application for solving solubility problems leads to essentially the same quality as for the original parameters. Based on provided models, it is possible to characterize any solvent and liquid solute for which HSP data are unavailable.

## 1. Introduction

Modeling of physicochemical properties of multicomponent systems, as, for example, solubility and miscibility, requires information about the nature of interactions between the components. A comprehensive and general characteristics of intermolecular interactions was introduced in 1936 by Hildebrandt [1]. This approach is based on the analysis of solubility parameters $\delta$ defined as the square root of the cohesive energy density, which can be estimated directly from enthalpy of vaporization, $\Delta H_v$, and molar volume (Eq. (1)):

$$\delta = \sqrt{\frac{\Delta H_v - RT}{V_m}}. \qquad (1)$$

Since the cohesive energy is the energy amount necessary for releasing the molecules' volume unit from its surroundings, the solubility parameter can be used as a measure of the affinity between compounds in solution. In his historical doctoral thesis [2], Hansen presented a concept of decomposition of the solubility parameter into dispersion (*d*), polarity (*p*), and hydrogen bonding (HB) parts, which enables a much better description of intermolecular interactions and broad usability [3, 4]. By calculating the Euclidean distance between two points in the Hansen space, one can evaluate the miscibility of two substances according to the commonly known rule "*similia similibus solvuntur.*" There are many scientific and industrial fields of Hansen solubility parameters application, including polymer materials, paints, and coatings (e.g., miscibility and solubility [5–9], environmental stress cracking [10, 11], adhesion [12], plasticizers compatibility [13], swelling, solvent diffusion, and permeation [14, 15], and polymer sensors designing [16], pigments and nanomaterials dispersibility [3, 17–20]), membrane filtration techniques [21], and pharmaceutics and pharmaceutical technology (e.g., solubility [22–27], cocrystal screening [28, 29], drug-DNA interaction [30], drug's absorption site prediction [31], skin permeation [32], drug-nail affinity [33], drug-polymer miscibility, and hot-melt extrusion technology [34–37]).

Due to the high usability of HSP, many experimental and theoretical methods of determining these parameters were proposed. For example, HSP can be calculated utilizing the equation of state [38] derived from statistical thermodynamics. Alternatively, models taking advantage of the additivity concept, such as the group contribution method (GC) [25, 39–41] is probably the most popular one. Despite the simplicity and success of these approaches, there are some important limitations. First of all, the definition of groups is ambiguous which leads to different parameterization provided by different authors [39]. Besides, the same formal group type can have varying properties, depending on the neighborhood and intramolecular context. As an alternative, molecular dynamics simulations were used for HSP values determination [16, 42–44] even in such complex systems as polymers. Interestingly, quantum-chemical computations were rarely used for predicting HSP parameters. However, the method combining COSMO-RS sigma moments and artificial neural networks (ANN) methodology [45] deserves special attention. Noteworthy, much better results were obtained using ANN than using the linear combination of sigma moments [45].

The application of nonlinear models is a promising way of HSP modeling. In recent times, there has been a significant growth of interest in developing QSPR/QSAR models utilizing nonlinear methodologies, like support vector machine [46–50] and ANN [51–55] algorithms. The attractiveness of these methods lies in their universality and accuracy. However, many are characterized by complex architectures and nonanalytical solutions. An interesting exception is the multivariate adaptive regression splines (MARSplines) [56]. This method has been applied for solving several QSPR and QSAR problems including crystallinity [57], inhibitory activity [58, 59], antitumor activity [60], antiplasmodial activity [61], retention indices [62], bioconcentration factors [63], or blood-brain barrier passage [64]. Interestingly, some studies suggested a higher accuracy of MARSplines when compared to ANN [57, 58, 65]. An interesting approach is the combination of MARSplines with other regression methods. As shown in the research on blood-brain barrier passage modeling, the combination of MARSplines and stepwise partial least squares (PLS) or multiple linear regression (MLR) gave better results than pure models [64]. The MARSplines model for a dependent (outcome) variable $y$ and $M + 1$ terms (including intercept) can be summarized by the following equation:

$$y = F_0 + \sum_{m=1}^{M} F_m \cdot H_{km}(x_{v(k,m)}), \quad (2)$$

where summation is over $M$ terms in the model, while $F_0$ and $F_m$ are the model parameters. The input variables of the model are the predictors $x_{v(k,m)}$ (the $k$th predictor of the $m$th product). The function $H$ is defined as a product of basis functions ($h$):

$$H_{ki}(x_{v(k,m)}) = \prod_{k=1}^{K} h_{km}(x_{v(k,m)}), \quad (3)$$

where $x$ represents two-sided truncated functions of the predictors at point termed knots. This point splits distinct regions for which one of the formula is taken, $(t - x)$ or $(x - t)$; otherwise, the respective function is set to zero. The values of knots are determined from the modeled data.

Since nonparametric models are usually adaptive and with a high degree of flexibility, they can very often result in overfitting of the problem. This can lead to poor performance of new observations, even in the case of excellent predictions of the training data. Such inherent lack of generalizations is also characteristic for the MARSplines approach. Hence, additionally to the pruning technique used for limiting the complexity of the obtained model by reducing the number of basis functions, it is also necessary to augment the analysis with the physical meaning of obtained solutions.

The purpose of this study is to test the applicability of the MARSplines approach for determining Hansen solubility parameters and to verify the usefulness of the obtained models by solubility predictions. Hence, an in-depth exploration was performed, including resizing of the models combined with a normalization and orthogonalization of both factors and descriptors. Also, a comparison with the traditional multivariable regression QSPR approach was undertaken. Finally, the obtained models were used for solving typical tasks for which Hansen solubility parameters can be applied, in order to document their reliability and applicability.

## 2. Methods

*2.1. Data Set and Descriptors.* In this paper, the data set of experimental HSP collected by Járvás et al. [45] was used for QSPR models generation. This diverse collection comprises a wide range of nonpolar, polar, and ionic compounds including hydrocarbons (e.g., hexane, benzene, toluene, and styrene), alcohols (e.g., methanol, 2-methyl-2-propanol, glycerol, sorbitol, and benzylalcohol), aldehydes and ketones (e.g., benzaldehyde, butanone, methylisoamylketone, and diisobutylketone), carboxylic acids (e.g., acetic acid, acrylic acid, benzoic acid, and citric acid), esters (isoamyl acetate, propylene carbonate, and butyl lactate), amides (N,N-dimethylformamide, formamide, and niacinamide), halogenated hydrocarbons (e.g., dichloromethane, 1-chlorobutane, chlorobenzene, 1-bromonaphthalene), ionic liquids, and salts (e.g., [bmim]PF6, [bmim]Cl, sodium salts of benzoic acid, *p*-aminobenzoic acid, and diclofenac). These data were obtained from the original HSP database [39, 66] and several other reports [67, 68]. After removing the repeating cases from the original collection, a set of 130 compounds, for which experimental data of HSP are available, was used.

Using information encoded in canonical SMILES, PaDEL software [69] offers 1444 descriptors of both 1D and 2D types. Not all of them can be used in modeling, and those descriptors which are not computable for all compounds or with zero variance were rejected from further analysis. The remaining 886 parameters were used for models definition.

*2.2. Computational Protocol.* Model building was conducted using absolute values of descriptors or orthogonalized data. Since there are different criteria for selecting independent variables from the pool of mutually related ones, two specific criteria were applied. The first one relied on the direct correlation with modeled HSP data if $R^2 > 0.01$. The second one used ranking offered by Statistica [70], tailored for regression analysis. These parameters were considered as nonorthogonal ones for which the Spearman correlation coefficient was higher than 0.7 ($R^2 > 0.49$). These different methods of orthogonalization led to different sets of descriptors used during application of QSPR or MARSplines approaches. Types of performed computations are summarized on Scheme 1.

*2.3. QSPR Approach.* The development of QSPR models and internal validation of the multiple linear regression (MLR) approach was conducted using QSARINS software 2.2.2 [71, 72]. The genetic algorithm (GA) for variable selection was applied during the generation of the models, which were defined with no more than 20 variables. The following fitting quality parameters were used for the model evaluation: determination coefficient ($R^2$), adjusted determination coefficient ($R_{adj}$)$^2$, Friedman's "lack of fit" (LOF) measure, global correlation among descriptors ($K_{xx}$) [73, 74], root-mean-square error, and mean absolute error (RMSE$_{tr}$ and MAE$_{tr}$) calculated for the training set and F (Fisher ratio). Also, the following internal validation parameters were used: leave-one-out validation measure ($Q_{loo}$)$^2$, cross-validation root-mean-square error, and mean absolute error (RMSE$_{cv}$ and MAE$_{cv}$).

## 3. Results and Discussion

Since the aim of this paper is the verification of the efficiency of predicting Hansen solubility parameters based on models derived using the MARSplines approach, two alternative procedures were adopted. The first one relies directly on the solution coming from application of the MARSplines procedure. The resulting factors were then used for assessment of $p$, $d$, and HB parameters. Alternatively, in the second step, the obtained factors were used as new types of descriptors and applied in the standard QSPR modeling along with the ones obtained from PaDEL. The premise of such attempt relied on the assumption that new factors, accounting for nonlinear contributions, combined with descriptors raise the accuracy of the model. The consistency of the models was checked using an internal validation procedure and additionally by applying them for solving some typical tasks that utilize Hansen solubility parameters. Particularly, the classification of polymers as soluble and nonsoluble ones in a set of solvents was compared with the original values of Hansen parameters. Similarly, the prediction of preferential solubility of some drugs was tested.

*3.1. MARSplines Models.* Several models were computed using the whole set of 886 available descriptors (run1 and run2). Typically, the size of the problem was restricted to 25 or 30 basis functions with the number of interactions increasing from 2 up to 10. For example, the simplest model restricted to 25 basis functions with no more than double interactions is denoted as (25, 2). For each model, the regressions were analyzed in two manners. Firstly, the direct application of the set of factors obtained from MARSplines was performed for solving regression equations. Since some of the generated factors have shown an apparent linear correlation, the orthogonalization of the factors was undertaken according to the two mentioned approaches. This resulted in two alternative models, usually of lower complexity.

*3.2. MARSplines Modeling of Parameter d.* Hansen solubility parameter $d$ is the measure of interaction energy via dispersion forces. As other contributions to Hansen solubility space, it is expressed as the density of cohesive energy. Among all three descriptors, this one seems to be the most difficult to predict. Fortunately, the MARSplines procedure performed quite well even in this case. The details of all developed models are provided in Figure 1, which offers several interesting conclusions. First of all, the models with satisfactory descriptive potential are quite complex, requiring several factors. Fortunately, the actual number of descriptors is usually much lower since many factors utilize the same molecular descriptors. Besides, models relying on the absolute values of descriptors outperform models constructed using normalized descriptors. This seems to be surprising since normalization should not lead to any change in the model quality; however, in the case of MARSplines, there is a significant gain in using absolute values. This can be attributed to the very nature of MARSplines, which is strictly a data-driven nonparametric procedure. Another interesting conclusion comes from inspection of trends indicated by the solid black lines. The rise of the number of interactions does not seriously improve the quality of predictions. Although the $d(30, 10)$ model is slightly better than $d(25, 2)$, it comes at a cost of additional three factors. This is a fortunate circumstance, suggesting that developing simpler models can be quite sufficient. In the case of the $d(25, 2)$ model, the value of the adjusted correlation coefficients $(R_{adj})^2$ is as high as 0.94. The formal mathematical formula of the MARSplines-derived model is analogical to a typical QSPR equation, although instead of descriptors, the MARSplines factors are present. In the case of the $d(25, 2)$ model, Eq. (4) defines the mathematical formula for computation of the $d$ parameter.

| | |
|---|---|
| run1* | Without both orthogonalization and normalization |
| run2* | Without orthogonalization but with normalization |
| run3 | Without normalization but with orthogonalization** separately for each parameter |
| run4 | With both normalization and orthogonalization** separately for each parameter |

*In this modeling, the whole set of available parameters was used (886 descriptors) for each parameter.

**Two rankings of descriptors were used. First one (A) was done according to direct correlation with modeled data that provided $R^2 > 0.01$. Application of first type of orthogonalization and exclusion of the parameters with $R^2 < 0.01$ reduced the number of descriptors down to 127 in the case of the $d$ parameter, 134 for $p$ parameter and 128 the most appropriate for the HB parameter. The second one (B) used ranking offered by Statistica, selecting the most suitable parameters for regression analysis. Application of the first type of orthogonalization and excluding parameters with $R^2 < 0.01$ reduced the number of descriptors down to 118 in the case of $d$ and HB parameters, and down to 124 for the $p$ parameter.

Scheme 1: Summary of MARSplines and QSPR runs.

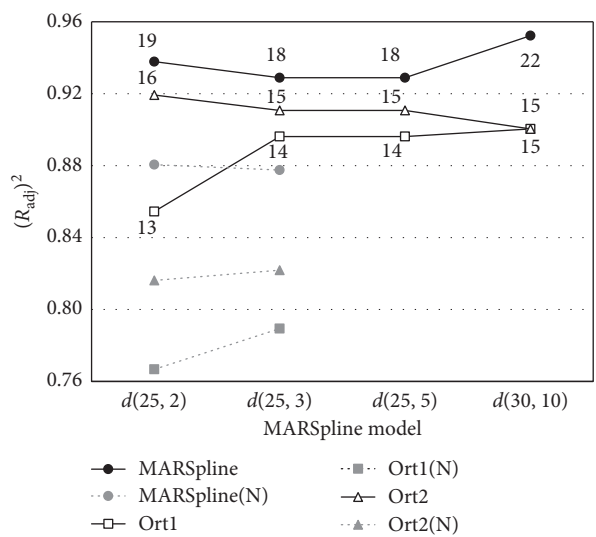

Figure 1: Results of predicting the values of the $d$ descriptor, based on a series of $d(b, i)$ MARSplines models characterized by number of initial basis functions ($b$) and allowed maximum interactions ($i$). Provided numbers represent amounts of factors used in the final regression function with statistically significant contributions. Grey lines represent results obtained after normalization of each of the descriptor distributions, while black lines correspond to models built on absolute values of descriptors.

Factors definitions, along with their contributions, were summarized in Table 1:

$$d(25, 2) = F_0 + \sum_{i=1}^{19} a_i \cdot F_i. \quad (4)$$

The values of coefficients come from the internal validation procedure performed using the QSARINS default algorithm. It is a typical many-leave-out procedure rejecting 30% of the data. The correlation between experimental and computed values of the $d$ solubility parameter is plotted in Figure 2. Both data for $d(25, 2)$ and $d(30, 1)$ models were provided. It is quite visible that the gain of the extended model is not very impressive, and for further applications, the $d$ parameter will be computed according to model defined by Eq. (4). Although formally there are nineteen factors in this equation, some can actually be consolidated as one. For example, F1 appears in definitions of F3, F4, F17, and F18. It seems to be rational to consolidate them into one by extraction of F1 and redefining the factors by multiplication of the sum of the remaining parts by F1. This in fact does not change the size of the problem, which should be attributed to the number of descriptors used in definition of MARSplines factors rather than factors. In the case of Eq. (4), twelve PaDEL descriptors are used. The majority of them (ATSC1i, AATS2e, AATS2p, ATSC3p, AATSC6v, ATSC1v, ATS4m, and GATS6c) belongs to 2D autocorrelation descriptors [75]. One descriptor VE3_Dzi is of the Barysz matrix type [75]. Besides, atom-type electrotopological state 2D descriptors (SsOH and minHCsats) were also included in the model [76–78]. Finally, the values of the nHBDon_Lipinski descriptor are also used in the model, and this parameter represents simply the number of hydrogen bond donors.

As it was mentioned beforehand, the construction of the models using MARSplines factors can in some cases lead to apparent mutual linear correlation between these factors. In all observed cases, these dependencies were really superficial and resulted from the fact that the basis functions used knots for splitting values below and above the given threshold. In such situation, the correlation, even if mathematically detectable, has no significant meaning and is artificial. From the formal point of view, it is possible to rearrange such factors in the regression function, consolidating them into one and removing these apparent correlations. However, it was interesting to observe if it is possible to reduce the number of factors in the model by eliminating these apparently nonorthogonal ones. For this purpose, two types of orthogonalization were performed, and the results are presented in Figure 2. First of all, the models were significantly worse compared to the original ones. This is not surprising, since after orthogonalization, fewer factors were used in the final regression function, which resulted not only from elimination of apparently related ones but also from the fact that correlation coefficients in new regressions were not statistically significant. Indeed, the reduction of the $d(25, 2)$ model by orthogonalization based on Statistica ranking led to a model with 16 factors and corresponding $(R_{adj})^2 = 0.92$.

TABLE 1: Regression factors along with their weights defining the $d(25, 2)$ MARSplines model in Eq. (4).

| Factor | $a_i \pm$ SD | Mathematical relationships |
|---|---|---|
| F0 | $16.6638 \pm 0.1485$ | |
| F1 | $0.0092 \pm 0.0015$ | max(0; ATSC1v + 144.0547) |
| F2 | $0.0648 \pm 0.0050$ | max(0; −6.51036-ATSC1i) |
| F3 | $-0.0002 \pm 0.0001$ | F1·max(0; SsOH-7.94125) |
| F4 | $0.0015 \pm 0.0001$ | F1·max(0; 7.94125-SsOH) |
| F5 | $1.5234 \pm 0.3405$ | max(0; AATS2e-7.54442) |
| F6 | $-3.4184 \pm 0.3990$ | max(0; 7.54442-AATS2e) |
| F7 | $-1.2270 \pm 0.2402$ | F5·max(0; minHCsats-4.17191) |
| F8 | $-6.0944 \pm 0.5530$ | F5·max(0; 4.17191-minHCsats) |
| F9 | $0.2519 \pm 0.0682$ | max(0; AATS2p-1.25641) |
| F10 | $-6.6966 \pm 1.3720$ | max(0; 1.25641-AATS2p) |
| F11 | $-0.0192 \pm 0.0036$ | max(0; ATS4m-2039.674)·F10 |
| F12 | $0.0021 \pm 0.0006$ | max(0; 2039.6739-ATS4m)·F10 |
| F13 | $1.5646 \pm 0.2463$ | max(0; nHBDon_Lipinski-2.00000)·F5 |
| F14 | $0.3218 \pm 0.1429$ | max(0; 2.00000-nHBDon_Lipinski)·F5 |
| F15 | $0.0208 \pm 0.0037$ | max(0; −144.0547-ATSC1v)·max(0; VE3_Dzi + 1.57191) |
| F16 | $-0.1155 \pm 0.0211$ | max(0; ATSC1i + 6.51036)·max(0; 1.00111-GATS6c) |
| F17 | $-0.0008 \pm 0.0002$ | F1·max(0; ATSC3p + 0.63792) |
| F18 | $-0.0031 \pm 0.0006$ | F1·max(0; −0.63792-ATSC3p) |
| F19 | $0.2626 \pm 0.0721$ | max(0; 0.00000-AATSC6v)·max(0; AATS2p-1.25641) |

Model statistics: fitting criteria: $N = 130$, $R^2 = 0.947$, $R^2_{adj} = 0.938$, $F = 103.39$, and LOF = 0.368; internal validation criteria: LMO (30%), $Q^2_{loo} = 0.860$, RMSE = 0.697, and MAE = 0.431.

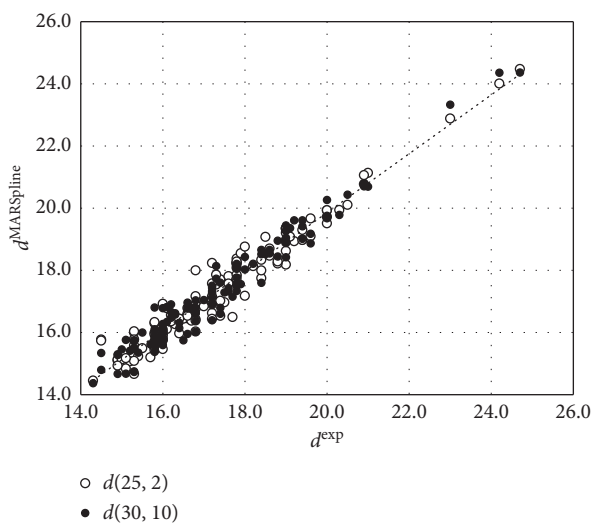

FIGURE 2: The correlation between experimental and computed values of parameter $d$ prediction is done using Eq. (1). The quality of the chosen optimal $d(25, 2)$ model is characterized by the fitting criteria: $R^2 = 0.9470$, $(R_{adj})^2 = 0.9378$, LOF = 0.3680, $K_{xx} = 0.4341$, RMSE$_{tr}$ = 0.4293, MAE$_{tr}$ = 0.3239, $F = 103.3872$, and $N = 130$, and fulfils the following internal validation criteria: $(Q_{loo})^2 = 0.8601$, RMSE$_{cv}$ = 0.6973, and MAE$_{cv}$ = 0.4309 [71, 72].

### 3.3. MARSplines Modeling of Parameter p.
Series of models for computing the polarity descriptor was also developed, and their predictive powers are summarized in Figure 3. The quality of the correlation between experimental values and the ones predicted using the best models is illustrated in Figure 4.

As one can infer from Figure 3, the best model with orthogonal factors is $p(30, 10)$. However, it is characterized by a high degree of descriptors interaction. Therefore, the

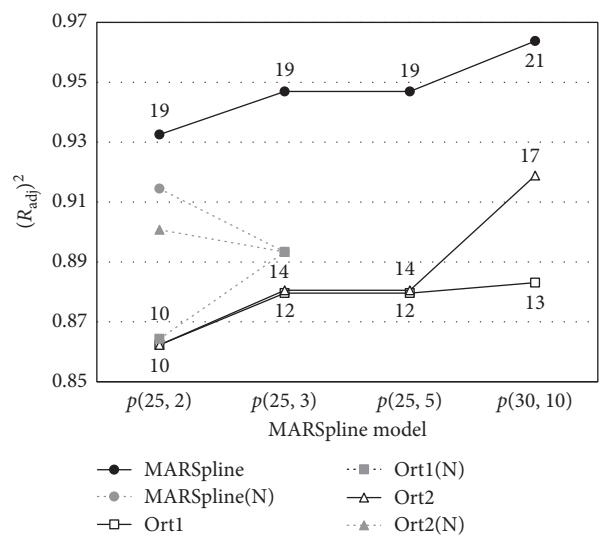

FIGURE 3: Results of predicting the values of the $p$ descriptor, based on a series of $p(b, i)$ MARSplines models. Notation is the same as in Figure 1.

most optimal one seems to be $p(25, 3)$. This model is expressed by Eq. (5), and the factors descriptions along with their contributions are summarized in Table 2. This model utilizes descriptors belonging to several classes, namely, information content (IC0 and ZMIC2) [75], autocorrelation (AATS2m, GATS1e, GATS2e, GATS5m, AATSC5i, ATSC5e, and MATS1v) [75], molecular linear-free energy relation (MLFER_S) [79], mindssC [76–78], and Petitjean topological and shape indices (PetitjeanNumber) [80]. The reduction of variables achieved using the genetic algorithm does not always guarantee that descriptors with clear meaning will be selected. Nevertheless, among descriptors

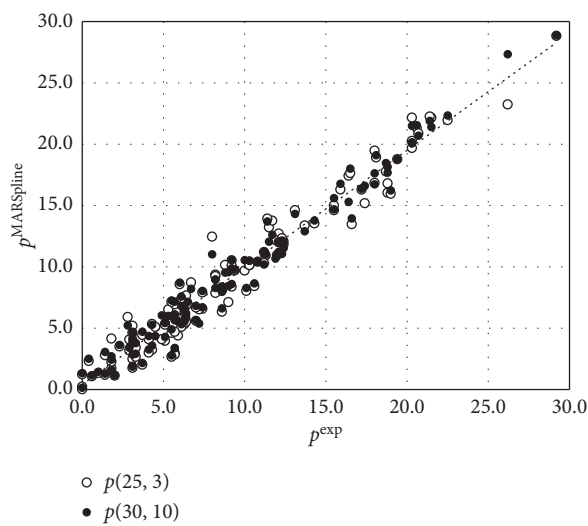

FIGURE 4: The correlation between experimental and computed values of parameter $p$ prediction is done using Eq. (1). The quality of the chosen optimal $p(25, 3)$ model is characterized by fitting criteria: $R^2 = 0.9425$, $(R_{adj})^2 = 0.9325$, LOF = 4.4911, $K_{xx} = 0.3758$, $RMSE_{tr} = 1.4998$, $MAE_{tr} = 1.1902$, $F = 94.8671$, and $N = 130$, and fulfils the following internal validation criteria: $(Q_{loo})^2 = 0.9100$, $RMSE_{cv} = 1.8765$, and $MAE_{cv} = 1.4655$ [71, 72].

TABLE 2: MARSplines $p(25, 3)$ model regression factors along with their weights.

| Factor | $a_i \pm SD$ | Mathematical relationships |
|---|---|---|
| F0 | $3.0017 \pm 0.2777$ | |
| F1 | $13.0874 \pm 1.3342$ | max(0; IC0-1.14332) |
| F2 | $-9.0702 \pm 2.4982$ | max(0; 1.14332-IC0) |
| F3 | $18.0918 \pm 1.7520$ | max(0; PetitjeanNumber-0.46154) |
| F4 | $-0.8421 \pm 0.2724$ | max(0; 60.09146-AATS2m)·F1 |
| F5 | $-25.2410 \pm 3.4481$ | max(0; 0.75379-GATS2e)·F1 |
| F6 | $51.6379 \pm 5.0897$ | F5·max(0; AATSC5i-0.48388) |
| F7 | $73.5427 \pm 8.2229$ | F5·max(0; 0.48388-AATSC5i) |
| F8 | $8.5172 \pm 0.8475$ | max(0; MLFER_S-0.54800) |
| F9 | $-0.1257 \pm 0.0262$ | max(0; ZMIC2-16.19833) |
| F10 | $0.7386 \pm 0.0940$ | max(0; 16.19833-ZMIC2) |
| F11 | $-20.5206 \pm 3.2197$ | F8·max(0; MATS1v + 0.17725) |
| F12 | $-16.5968 \pm 2.2740$ | F8·max(0; −0.17725-MATS1v) |
| F13 | $-28.6245 \pm 4.1609$ | max(0; GATS5m-0.54611)·max(0; GATS2e-0.75379)·F1 |
| F14 | $-48.3050 \pm 7.0216$ | max(0; 0.54611-GATS5m)·max(0; GATS2e-0.75379)·F1 |
| F15 | $67.3423 \pm 17.0712$ | max(0; −0.26841-ATSC5e)·max(0; 0.46154-PetitjeanNumber) |
| F16 | $4.7141 \pm 1.0570$ | max(0; 60.09146-AATS2m)·max(0; mindssC + 0.24537)·F1 |
| F17 | $2.0457 \pm 0.4563$ | max(0; 60.09146-AATS2m)·max(0; −0.24537-mindssC)·F1 |
| F18 | $82.5944 \pm 16.2082$ | max(0; GATS2e-0.75379)·max(0; GATS1e-0.84779)·F1 |
| F19 | $116.1381 \pm 25.4572$ | max(0; GATS2e-0.75379)·max(0; 0.84779-GATS1e)·F1 |

Model statistics: fitting criteria: $N = 130$, $R^2 = 0.954$, $R^2_{adj} = 0.945$, $F = 122.2$, and LOF = 3.533; internal validation criteria: LMO (30%), $Q^2_{loo} = 0.935$, RMSE = 1.771, and MAE = 1.247.

which appeared in the $p(23, 3)$ model, IC0 and MLFER_S are quite simple to interpret in the context of polarity HSP since IC0 index expresses the diversity (heterogeneity) of atomic types [81], while MLFER_S is associated with the dipolarity/polarizability features of molecules [57, 82, 83]. Also autocorrelation descriptors GATS1e, GATS2e, and MATS1v deserve for special attention. In general, autocorrelation indices do not have a clear interpretation. Nevertheless, their appearance seems to be understandable since these descriptors were applied in different solubility prediction models reported previously [84–86]:

$$p(25, 3) = F_0 + \sum_{i=1}^{19} a_i \cdot F_i. \quad (5)$$

### 3.4. MARSplines Modeling of Parameter HB.

Analogously to the previously discussed parameters, the model corresponding to the hydrogen bonds interactions was developed and optimized. The results are summarized in Figures 5 and 6.

As it can be observed in the abovementioned figures, the HB(25, 2) model is characterized by the highest correlation between experimental and predicted values, comparing to previously discussed $d(25, 2)$ and $p(25, 3)$ models. The regression equation of HB(25, 2), along with factors descriptions, is defined as follows (Eq. (6); Table 3):

$$p(25, 2) = F_0 + \sum_{i=1}^{22} a_i \cdot F_i. \quad (6)$$

The HB(25, 2) model consists of 22 factors. However, it turned out, based on the QSPR methodology, that two of them (F4 and F5) have a zero contribution. The factors in the HB(25, 2) model were generated using the following descriptors: atom-type electrotopological state (SHBd) [76–78], information content (SIC1) [75], autocorrelation (GATS2e, AATSC1i, AATSC2i, and ATSC1v) [75], eccentric connectivity (ECCEN) [87], extended topochemical (ETA_dEpsilon_D) [88, 89], weighted path (WTPT-4) [90], Barysz matrix-based (VE3_DzZ) [75], and Crippen's (CrippenLogP) parameters [91]. Noteworthy, SHBd, ETA_dEpsilon_D, and CrippenLogP molecular descriptors that appeared in the above model are quite intuitive in the context of HB parameter interpretation. The SHBd descriptor is simply the sum of all E-States corresponding to hydrogen bonds donors [76–78]. ETA_dEpsilon_D parameter is also associated with hydrogen bonds donating abilities. Thus, both SHBd and ETA_dEpsilon_D descriptors have been used for QSAR protein binding/inhibition problems solving [92–95]. The appearance of CrippenLogP, being a part of the F3 factor, is understandable since more polar molecules are usually more likely to form strong hydrogen bonds. Noteworthy, LogP, which is probably one of the most popular polarity parameters, was used for the Yalkowsky model [96, 97], which confirms its usability in the HSP approach. Based on the F3 definition (Table 3), an interesting observation can be made; when CrippenLogP values are lower than about −2.34, the polarity is extremely high and so it does not affect the ability to form hydrogen bonds. This treatment of variables, associated with the determination of their scope of application, is characteristic for the MARSplines methodology. Similarly, as in case of other HSP models, autocorrelation descriptors play an important role. These molecular measures are related to the basic atomic properties such as Sanderson electronegativities (GATS2e), ionization potential (AATSC1i and AATSC2i), and van der Waals volume (ATSC1v).

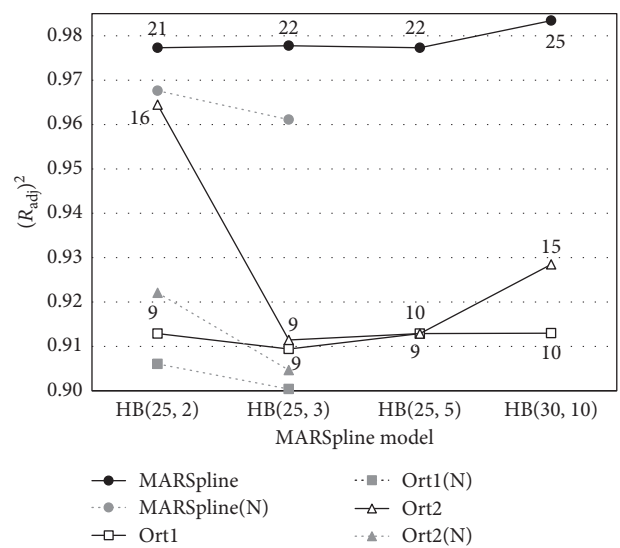

Figure 5: Results of predicting the values of the HB descriptor, based on a series of HB($b, i$) MARSplines models. Notation is the same as in Figure 1.

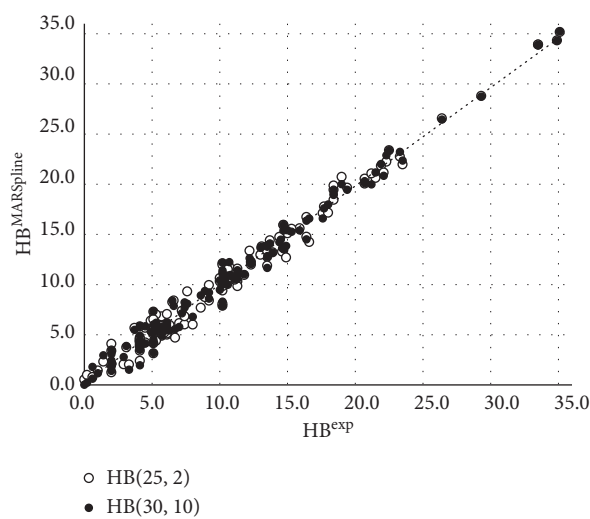

Figure 6: The correlation between experimental and computed values of parameter HB. Prediction is done using Eq. (1). The quality of the chosen optimal HB(25,2) model is characterized by the fitting criteria: $R^2 = 0.9812$, $(R_{adj})^2 = 0.9773$, LOF = 2.4449, $K_{xx} = 0.4654$, RMSE$_{tr}$ = 1.0344, MAE$_{tr}$ = 0.8222, $F$ = 253.5683, and $N$ = 130, and fulfils the following internal validation criteria: $(Q_{loo})^2 = 0.9670$, RMSE$_{cv}$ = 1.3696, and MAE$_{cv}$ = 1.0381 [71, 72].

### 3.5. QSPR Models.

QSARINS software [71, 72] offers a straightforward method for regression analysis, especially efficient in the case of large QSPR problems. In such cases, the complete exploration of all possible combinations of descriptors is prohibited by too large numbers of potential arrangements of the variables. In such situation, the genetic algorithm [98] offers a rational way of exploration of the most promising regions of QSPR solution space. Here, all QSPR models were built based on orthogonal sets of descriptors, that is denoted as run3 and run4, according to two different ways of orthogonalization (Scheme 1). Besides,

Table 3: MARSplines HB(25, 2) model regression factors along with their weights.

| Factor | $a_i \pm$ SD | Mathematical relationships |
|---|---|---|
| F0 | $12.6280 \pm 0.4535$ | |
| F1 | $5.5560 \pm 0.7079$ | max(0; SHBd-0.84757) |
| F2 | $-10.4070 \pm 1.0496$ | max(0; 0.84757-SHBd) |
| F3 | $1.0900 \pm 0.1333$ | max(0; 2.3406-CrippenLogP) |
| F4 | $0.0000 \pm 0.0000$ | max(0; ECCEN-20.00000)·F2 |
| F5 | $0.0000 \pm 0.0000$ | max(0; 20.00000-ECCEN)·F2 |
| F6 | $-4.0810 \pm 0.4901$ | max(0; GATS2e-0.92565) |
| F7 | $-4.9500 \pm 0.5455$ | max(0; 0.92565-GATS2e) |
| F8 | $-0.1460 \pm 0.0470$ | max(0; WTPT-4-2.32775) |
| F9 | $-1.5640 \pm 0.1466$ | max(0; 2.32775-WTPT-4) |
| F10 | $-62.8000 \pm 7.3785$ | F1·max(0; SIC1-0.59306) |
| F11 | $-20.6450 \pm 5.3855$ | F1·max(0; 0.59306-SIC1) |
| F12 | $21.0280 \pm 3.0488$ | max(0; ETA_dEpsilon_D-0.05394) |
| F13 | $79.3130 \pm 14.5139$ | max(0; 0.05394-ETA_dEpsilon_D) |
| F14 | $-0.3920 \pm 0.0593$ | max(0; VE3_DzZ + 3.00162)·F8 |
| F15 | $-88.4270 \pm 13.1857$ | max(0; AATSC1i + 0.83463)·F13 |
| F16 | $-100.3560 \pm 19.2748$ | max(0; $-0.83463$-AATSC1i)·F13 |
| F17 | $3.4670 \pm 0.5511$ | max(0; AATSC7i-0.42042) |
| F18 | $3.1050 \pm 0.6674$ | max(0; 0.42042-AATSC7i) |
| F19 | $0.1370 \pm 0.0591$ | max(0; ATSC1v + 23.64635)·F12 |
| F20 | $0.2160 \pm 0.0470$ | max(0; $-23.64635$-ATSC1v)·F12 |
| F21 | $1.8170 \pm 0.7239$ | F2·max(0; AATSC2i + 0.09514) |
| F22 | $6.9340 \pm 1.5981$ | F2·max(0; $-0.09514$-AATSC2i) |

Model statistics: fitting criteria: $N = 130$, $R^2 = 0.974$, $R^2_{adj} = 0.970$, $F = 216.6$, and LOF = 2.955; internal validation criteria: LMO (30%), $Q^2_{loo} = 0.960$, RMSE = 1.509, and MAE = 1.150.

additional QSPR runs were performed with factors augmenting the pool of descriptors. Orthogonalization was performed within the extended set of descriptors favoring MASRpline factors, which ensured that factors were not directly correlated with original descriptors, what is of course possible. The results of these series of computations are presented in Figures 7–9.

The results of computing the dispersion parameter are provided in Figure 7. The developed models are of varying size, starting from 2 up to 20 parameters. However, QSPR models are fairly saturated starting from nine parameters. The most important message coming from Figure 7 is that the classical QSPR formalism leads to modes which are significantly less accurate compared to MARSplines. Even models with several parameters do not reach the quality of description offered by the model defined by Eq. (4). Inclusion of all MARSplines factors into the pool of descriptors leads to a serious improvement of linear regression approach but is still far from the best solution. It seems that, in the case of the $d$ parameter, there is no gain in combination of MARSplines factors with PaDEL descriptors and searching for the solution via the QSPR approach. Similar conclusion can be drawn based on plots provided in Figure 8, documenting the accuracy of the models developed for computing the $p$ parameter. However, since in this case, there is a serious discrepancy between the original MARSplines model and the reduced one, and some QSPR models exceed the accuracy of the latter. Only 20-parameter regression functions reach similar accuracy as the MARSplines model defined by Eq. (5). Finally, similar analysis was performed for modeling of the HB parameter. This time a quite different set of data was obtained, as documented in Figure 9. Quite satisfying accuracy can be achieved even when 4 factors are used in the QSPR equation. Besides, there is a much steeper growth of the $(R_{adj})^2$ parameter compared to $d$ and $p$ HSP models, which are less sensitive to the pool of descriptors. Also, in the case of HB parameter, the solution obtained by application of the MARSplines approach offers the highest accuracy.

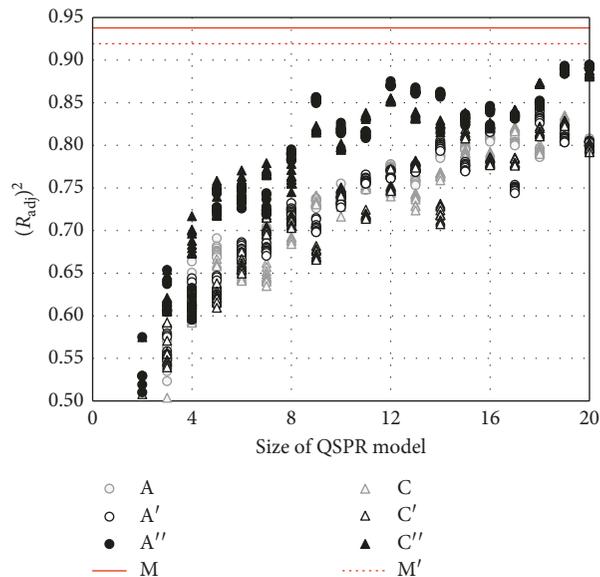

Figure 7: Distributions of $(R_{adj})^2$ values characterizing a variety of QSPR models predicting $d$ parameter based on PaDEL descriptors or factors resulting from MARSplines models. Open grey symbols represent models built using unnormalized parameters orthogonalized in two ways. Open black symbols stand for similar models but with normalized data. Filled black symbols denote QSPR models obtained by augmenting descriptors pool with orthogonal MARSplines factors. Red line documents the quality of the model obtained using all factors identified in the MARSplines procedure (Eq. (4)).

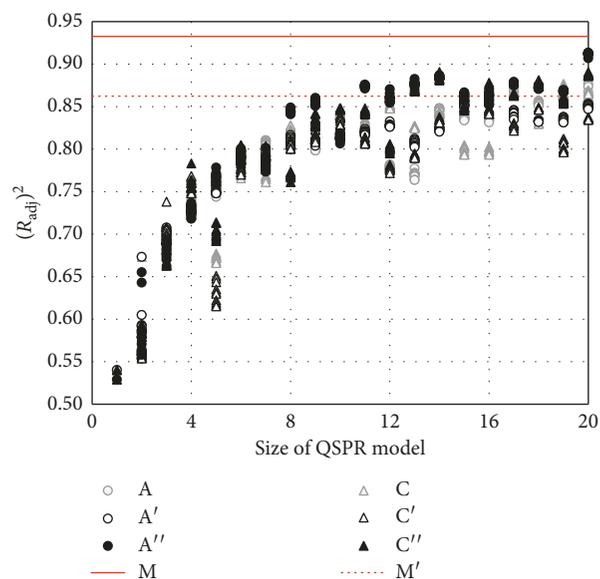

Figure 8: Distributions of $(R_{adj})^2$ values characterizing a variety of QSPR models predicting the $p$ Hansen solubility parameter based on PaDEL descriptors or factors resulting from MARSplines models. Notation is the same as in Figure 7.

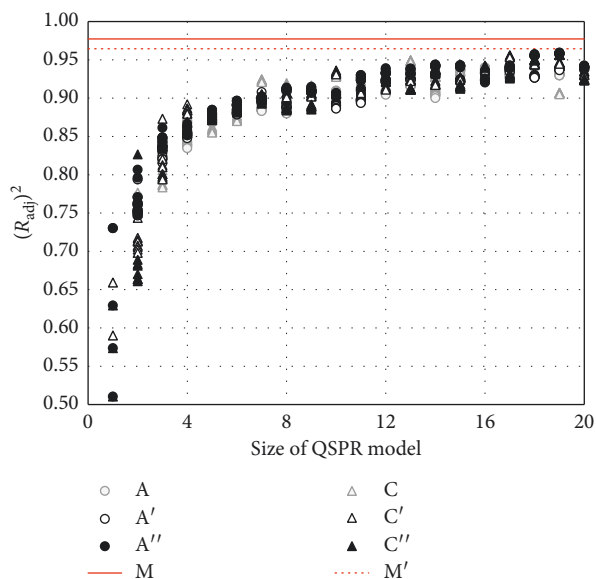

Figure 9: Distributions of $(R_{adj})^2$ values characterizing a variety of QSPR models predicting the HB Hansen solubility parameter based on PaDEL descriptors or factors resulting from MARSplines models. Notation is the same as in Figure 7.

3.6. Applications of MARSplines Models. One of the most often used and direct applications of Hansen solubility parameters is the selection of appropriate solvent for solubilization or dispergation of different solids and materials including drugs [22–26], polymers [5–9], herbicides [7], pigments and dyes [3, 18], and biomaterials [99]. It is typically done by computing HSP parameters based on a series of solubility measurements. Typically, 20–30 solvents are used for covering a broad range of Hansen parameters space [20, 39, 100, 101]. Alternatively, mixtures of two solvents are prepared in such a way that the broad range of HSP is covered by solutions [102–106]. The formal procedure of solvents classification utilizes some threshold of solubility for distinguishing soluble cases from nonsoluble ones. Different criteria may be applied, but very often, the dissolution of the solid solute below 1 mg per 100 ml is considered as insoluble [107–110]. Hence, the solubility measurements can be reduced to the list of good and bad solvents, which resembles strong or weak interactions of the tested media with considered substance or material. The collection of three HSP parameters for all the solvents is plotted in a 3-dimensional space providing the location of solubility spheres. Additionally, empirical parameter defining the size of the sphere is computed for maximizing the classification for highest prediction rate of experimentally derived binary solubility data. This minimization protocol can be done using dedicated software, as, for example, HSPiP (Hansen solubility parameters in practice) [66]. However, it is also possible to take advantage of the definition of the contingency table or confusion matrix often used to describe the performance of a classification model. Here, this strategy was adopted for the solubility classification by using the straightforward procedure of maximizing the values of balanced accuracy (BACC = (TP/P + TN/N)/2), where TP and TN denote true positives and negatives, while P and N represent all positive and negative cases, respectively. This measure is one of the most commonly used ways of quantification of binary classifiers. It seems to be a natural adaptation of this terminology for rating the solubility as a mathematically coherent approach. Besides, no dedicated software is necessary, and any solver-like algorithms can be applied. The results provided below were computed using the evolutionary algorithm implemented in Excel.

3.7. Application of HSP Models to Polymers Dissolution. The collection of the polymer solubility data was taken from the literature [39]. The experimentally measured data were originally classified on a scale described by the following qualifiers: (1) soluble, (2) almost soluble, (3) strongly swollen and slight solubility, (4) swollen, (5) little swelling, and (6) no visible effect. This list was converted into binary data by assuming polymer solubility only in the first case and treating other situations as nonsoluble polymers. For the whole set of 33 polymers for which solubility was determined in 85 solvents, the classification was done by optimization of all three HSP, as well as $R_o$ for each polymer. The solubility was predicted based on the classical formula of the distance in HSP space as follows:

$$R = \sqrt{4\left(\delta_d^P - \delta_d^S\right)^2 + \left(\delta_p^P - \delta_p^S\right)^2 + \left(\delta_h^P - \delta_h^S\right)^2}, \quad (7)$$

where the subscript P denotes the polymer and S the solvent. Four sets of solvent parameters were tested. They corresponded to (a) our model provided this paper in Eqs. (4)–(6), (b) original set of parameters collected in Table A1 of "Hansen solubility parameters: a user's handbook. Appendix A" [39], (c) collection provided by Járvás et. al [45], and (d) HSP parameters from the green solvent set [111]. Following the Hansen concept, the relative energy difference (RED) is defined by the following ratio:

$$\text{RED} = \frac{R}{R_0}, \quad (8)$$

where $R_0$ denotes the tolerance radius of a given polymer. In this approach, the material characterized by the model as RED > 1 is considered to be resistant to a solvent, whereas cases for which RED < 1 are regarded as soluble. During the procedure of solubility classification, the HSP values characterizing the solvent were kept intact and only the parameters for the polymer were adjusted for maximizing BAC for the whole set. The results of these computations are summarized in Table 4.

In all cases, the identification of true positive and true negative cases was higher than 90%. The misclassification of

Table 4: Results of the solubility classification of 33 polymers in 85 solvents [39].

| Data set* | TP | TN | FP | FN |
|---|---|---|---|---|
| A | 90.8% ± 7.2% ($p = 1.00$) | 91.6% ± 7.0% ($p = 1.00$) | 9.2% ± 7.2% ($p = 1.00$) | 8.4% ± 7.0% ($p = 1.00$) |
| B | 91.1% ± 6.9% ($p = 0.88$) | 92.4% ± 7.0% ($p = 0.66$) | 8.9% ± 6.9% ($p = 0.88$) | 7.6% ± 7.0% ($p = 0.66$) |
| C | 93.7% ± 5.7% ($p = 0.08$) | 92.1% ± 6.9% ($p = 0.80$) | 6.3% ± 5.7% ($p = 0.08$) | 7.9% ± 6.9% ($p = 0.80$) |
| D | 93.0% ± 6.0% ($p = 0.20$) | 92.3% ± 6.6% ($p = 0.70$) | 7.0% ± 6.0% ($p = 0.20$) | 7.7% ± 6.6% ($p = 0.70$) |

*A, MARSplines (25, 2) model; B, [39]; C, [45]; D, [111].

Table 5: Results of classification of API solubilities.

|  | [25] | This paper | TP (%) | TN (%) | BACC |
|---|---|---|---|---|---|
| Benzoic acid | 18 of 29 | 17 of 29 | 81.30 | 30.80 | 0.56 |
| Salicylic acid | 13 of 19 | 11 of 19 | 36.40 | 87.50 | 0.62 |
| Paracetamol | 14 of 24 | 18 of 24 | 50.00 | 92.90 | 0.71 |
| Aspirin | 14 of 23 | 14 of 23 | 46.20 | 80.00 | 0.63 |

soluble pairs as insoluble ones and vice versa was always lower than 10%. Although the results of classification using our models are somewhat worse, the difference is not statistically significant, and all approaches lead to the same quality of polymers solubility classification.

*3.8. Application of HSP Models to Drug-Like Solids Dissolution.* As the second type of external validation of the proposed model via application of the HSP procedure, the classification of solubility of drug-like solid substances was undertaken. Solubilities of benzoic acid, salicylic acid, paracetamol, and aspirin were taken from Stefanis and Panayiotou paper [25]. Again, maximizing of BACC was done by adopting HSP parameters. The results of the performed classification are collected in Table 5. In the third column of Table 5, there is provided the success rate obtained based on HSP values computed using the proposed model (Eqs. (4)–(6)), confronted with the success rate of the HSP approach adopted by Stefanis and Panayiotou [25] in the second column. It is worth mentioning that these authors used four parameters by splitting the hydrogen bonding part into donor and acceptor contributions. As it is documented in Table 5, the solubility predictions are almost of the same quality. In the case of benzoic acid and salicylic acid, a slightly lower quality of prediction was achieved. On the contrary, in the case of paracetamol, the success rate of the MARSplines model is higher.

The predictions based on the HSP, presented in the Tables 4 and 5, are characterized by quite good accuracy. However, it should be taken into account that, there are also other approaches which were successfully used for solubility prediction, classification, and ranking such as linear solvation energy relationship (LSER) models including the Abraham equation [112, 113] and the partial solvation parameters (PSPs) approach [114, 115], conductor-like screening model for real solvents (COSMO-RS) [116–118], UNIFAC [119–121], and finally (modified separation of cohesive energy density) MOSCED methodology [122, 123] which is an interesting extension of the HSP method. Nevertheless, HSP are, due to their universality, still very popular in solving many solubility and miscibility problems. In addition, it is also worth noting that, the proposed MARSplines model is characterized by a relatively high accuracy, although it was based only on the simplest 1D and 2D structural information retrieved from the SMILES code. Therefore, the model can be extended with more complex molecular descriptors, such as quantum-chemical indices.

## 4. Conclusions

MARSplines has been found to be a very effective way of generating factors suitable for prediction of three Hansen solubility parameters. The most important factor is preserving the formal linear relationship typical for QSPR studies and extending the model with nonlinear contributions. These come from the basis function definition and splitting the variable range into subdomains separated by knots values. Besides, factors used in the definition of the regression equations are constructed by multiplication of some number of basis functions that is referred to as the level of interactions. It is possible to formulate models with acceptable accuracy and user-defined complexity in terms of the number of basis functions and the level of interactions. It has been found that, for all three HSP parameters studied here ($p$, $d$, and HB), a promising precision was provided by quite simple models. The initial number of basis functions limited to 25 was found to be sufficient along with at most binary or ternary interaction levels. The internal validation of these models proved their applicability. The combination of descriptors with factors was also tested, but the obtained solutions were discouraging. Typical QSPR procedure relying on genetic algorithms for selecting the most adequate descriptors failed in finding models of the quality comparable with MARSplines. Only in the case of HB parameters, the result of the best QSPR models reached accuracy close to the MARSplines approach. Hence, it is not advised to combine traditional QSPR approaches by augmenting the pool of descriptors with factors derived in MARSplines. The observed supremacy of the latter in the case of HSP prediction suggests using it as a standalone procedure, especially since it offers a similar formal equation as traditional QSPR.

The application of the HSP models derived using MARSplines for typical solubility classification problems leads to essentially the same predictions as for the experimental sets of HSP. This conclusion is a promising circumstance for further development of multiple linear regression models augmented with nonlinear contributions.

## Data Availability

The data used to support the findings of this study are included within the article.

## Conflicts of Interest

## Acknowledgments

The provided free license of QSARINS by Prof. Paola Gramatica is warmly acknowledged. The research did not receive specific funding but was performed as part of the employment of the authors at Faculty of Pharmacy, Collegium Medicum of Bydgoszcz, Nicolaus Copernicus University in Toruń.

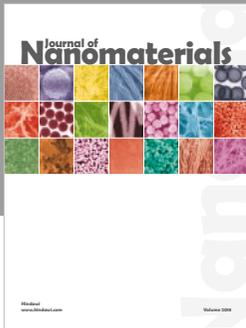
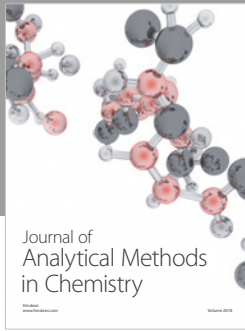
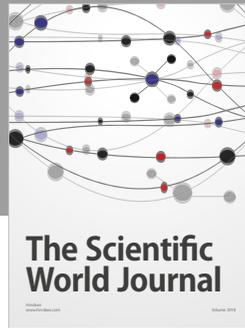
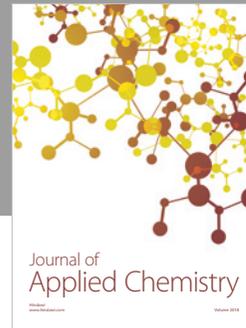
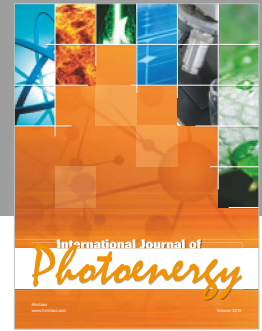
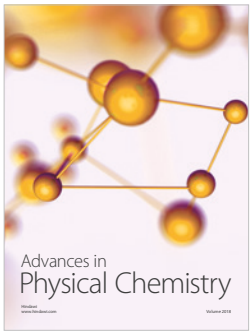
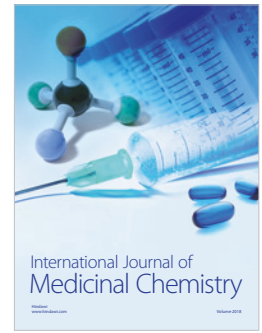
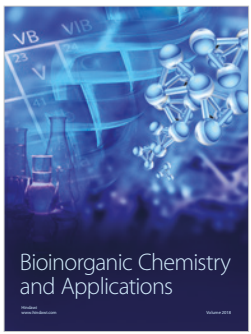
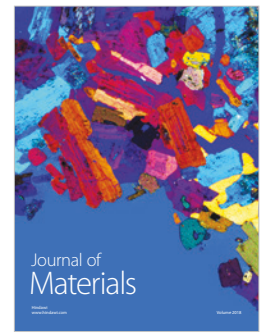
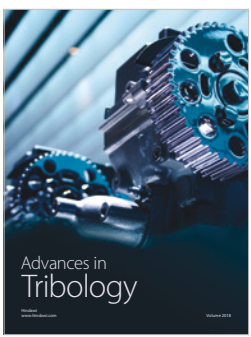
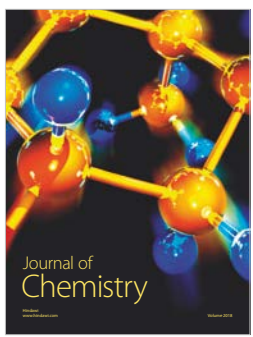
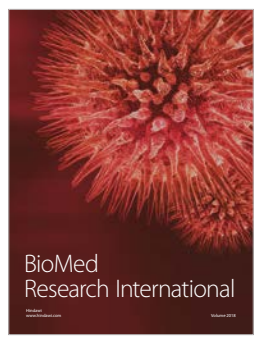
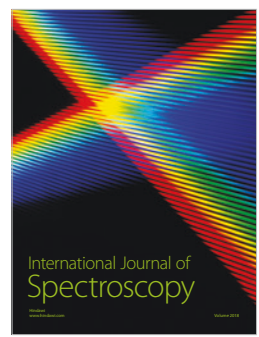
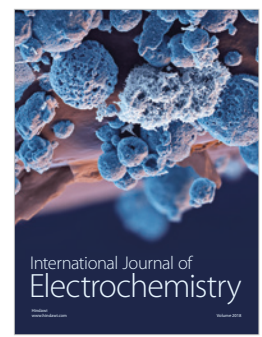
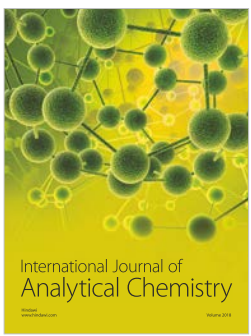
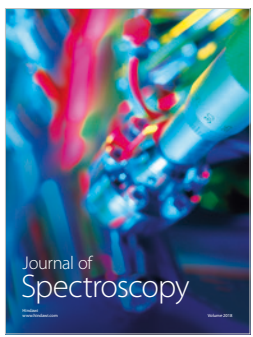
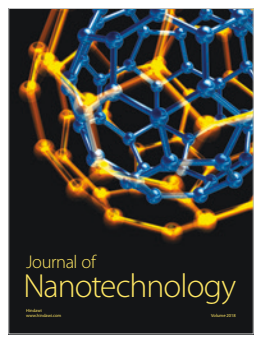
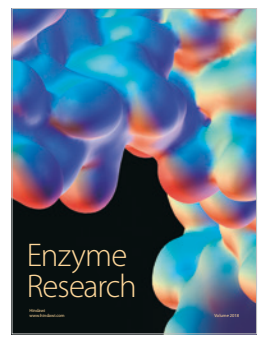
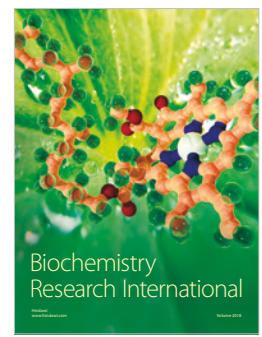

Submit your manuscripts at
www.hindawi.com